\documentclass[superscriptaddress,twocolumn,floatfix]{revtex4-1}
\usepackage{amssymb,amsmath,amsfonts,amsthm,dsfont}
\usepackage[normalem]{ulem}
\usepackage{bbm}
\usepackage{physics}
\usepackage{graphicx,wasysym}
\usepackage{xcolor}
\usepackage{float}
\usepackage{mathtools}
\usepackage[ruled,vlined]{algorithm2e}

\begin{document}
\title{Near Term Algorithms for Linear Systems of Equations}

\author{Aidan Pellow-Jarman}
\email{aidanpellow@gmail.com}
\affiliation{School of Mathematics, Statistics and Computer Science, University of KwaZulu-Natal, Durban, KwaZulu-Natal, 4001, South Africa}

\author{Ilya Sinayskiy}
\email{sinayskiy@ukzn.ac.za}
\affiliation{Quantum Research Group, School of Chemistry and Physics, University of KwaZulu-Natal, Durban, KwaZulu-Natal, 4001, South Africa}
\affiliation{National  Institute  for  Theoretical  and  Computational  Sciences  (NITheCS),  South  Africa}

\author{Anban  Pillay}
\email{pillayw4@ukzn.ac.za}
\affiliation{School of Mathematics, Statistics and Computer Science, University of KwaZulu-Natal, Durban, KwaZulu-Natal, 4001, South Africa}
\affiliation{Centre for Artificial Intelligence Research (CAIR), cair.org.za}

\author{Francesco  Petruccione}
\email{petruccione@ukzn.ac.za}
\affiliation{Quantum Research Group, School of Chemistry and Physics, University of KwaZulu-Natal, Durban, KwaZulu-Natal, 4001, South Africa}
\affiliation{National  Institute  for  Theoretical  and  Computational  Sciences  (NITheCS),  South  Africa}

\begin{abstract}
Finding solutions to systems of linear equations is a common prob\-lem in many
areas of science and engineering, with much potential for a speed up on quantum
devices. While the Harrow-Hassidim-Lloyd (HHL) quantum algorithm
yields up to an exponential speed up over classical algorithms in some cases, it requires a
fault tolerant quantum computer, which is unlikely to be available in the
near-term. Thus,  attention has turned to the investigation of quantum
algorithms for noisy intermediate-scale quantum (NISQ) devices where several
near-term approaches to solving systems of linear equations have been proposed. This
paper focuses on the Variational Quantum Linear Solvers (VQLS), and other closely related methods. This paper makes several contributions that include: the
first application of the Evolutionary Ansatz to the VQLS (EAVQLS), the first
implementation of the Logical Ansatz VQLS (LAVQLS), based on the Classical Combination of Quantum States (CQS) method , the first proof of principle demonstration of the CQS method on real quantum hardware and a method for the implementation of the Adiabatic Ansatz (AAVQLS). These approaches are implemented and contrasted. 
\end{abstract}

\maketitle

\section{Introduction}
\label{intro}
Systems of linear equations play an important role in many areas of science and engineering, making the potential quantum speed-up for solving them of great interest. Solving a system of $N$ linear equations with $N$ unknowns, expressible as $\mathbf{A}\vec{x}=\vec{b}$, involves finding the unknown solution vector $\vec{x}$ satisfying $\mathbf{A}\vec{x}=\vec{b}$. This is known as the Linear Systems Problem (LSP).

The Harrow-Hassidim-Lloyd (HHL) algorithm \cite{hhl2009} is a proposed quantum algo\-rithm for the quantum linear systems problem (QLSP) \cite{qlsp2018}, a quantum analogue of the LSP. The QLSP is stated as follows: Let $\mathbf{A}$ be an $N\times N$ hermitian matrix (however this algorithm is not limited to a hermitian matrix) and let $\vec{x}$ and $\vec{b}$ be $N$ dimensional vectors, satisfying $\mathbf{A}\vec{x}=\vec{b}$, having corresponding quantum states $\ket{x}$ and $\ket{b}$, such that
\begin{center}
\begin{gather}
     \ket{x}  \vcentcolon = \dfrac{\sum_{i}x_{i}\ket{i}}{\lvert\lvert \sum_{i}x_{i}\ket{i} \rvert\rvert_{2}}, \\
     \ket{b} \vcentcolon =  \dfrac{\sum_{i}b_{i}\ket{i}}{\lvert\lvert \sum_{i}b_{i}\ket{i} \rvert\rvert_{2}}.
\end{gather}
\end{center}
If \begin{math}\mathbf{A}\end{math} is not hermitian, define \begin{math}\Tilde{\mathbf{A}}=\bigl(\begin{smallmatrix}0&\mathbf{A}\\ \mathbf{A^{\dagger}}&0\end{smallmatrix}\bigr)\end{math}, which is hermitian, and instead solve the equation \begin{math}\Tilde{\mathbf{A}}\vec{y} = \bigl( \begin{smallmatrix}\vec{b} \\ 0 \end{smallmatrix} \bigr)\end{math} and solve for \begin{math}\vec{y} = \bigl( \begin{smallmatrix}0 \\ \vec{x} \end{smallmatrix} \bigr)\end{math}. Given access to matrix \begin{math}\mathbf{A}\end{math}\ by means of an oracle, and a unitary gate $\textit{U}$ such that $\textit{U}\ket{0} = \ket{b}$, output a quantum state $\ket{x'}$ such that $\lvert\lvert\ket{x}-\ket{x'}\rvert\rvert_{2}  \leq \epsilon$,  where $\epsilon$ is the error-bound of the approximate solution. 

The HHL algorithm is a quantum algorithms expected to give a substantial speed-up over classical approaches, providing up to an exponential speed-up over known classical algorithms in cases where the linear system is sparse, the condition number is low, and the actual solution vector is not required to be read out, but instead some scalar measure on the solution vector is of interest. As with many promising quantum algorithms, the HHL algorithm requires a fault-tolerant quantum computer to be successfully implemented, predicted to only be available in the long term future. 

Approaches at finding algorithms for noisy intermediate-scale quantum (NISQ) devices \cite{nisq}, availa\-ble in the near-term future, have focused mainly on a class of algorithms known as Variational Hybrid Quantum Classical Algorithms (VHQ\-CAs). The idea behind VHQCAs is to utilize a quantum-classical feedback loop. Here a quantum device is used to compute a cost function for a parameterized quantum circuit (ansatz), much more efficiently than is possible on a classical device\cite{vqls_1}, while a classical device is used to optimise the selection of the ansatz parameters. VHQCAs rely on the use of short depth quantum circuits to make them more resistant to noise and allowing them to be successfully run on NISQ quantum hardware. The main difficulties of this approach lie in  overcoming the noise inherent in quantum devices and the difficulty of optimizing the ansatz parameters. An example of these difficulties is the barren plateau problem \cite{barren}.

The Variational Quantum Eigensolver (VQE) \cite{vqe} is one notable VHQCA that solves the optimization problem,
\begin{center}
\label{equation_1}
\begin{gather}
     E = \min_{\theta} \bra{V(\mathbf{\theta})}H\ket{V(\mathbf{\theta})},
\end{gather}
\end{center}
whereby the minimum eigenvalue $E$ of some Hamiltonian $H$ is approximated through the optimization of $\mathbf{\theta}$ for some ansatz $V(\mathbf{\theta})$.

The Variational Quantum Linear Solver \cite{vqls_1, vqls_2} is based on the VQE, recently proposed to solve the quantum linear systems problem. Since its proposal, many variations have also been presented in order to overcome various difficult\-ies faced by the algorithm, and VHQCAs in general. Attempts to combat these difficult\-ies include training approaches like the Adiabatic Assisted VQE \cite{vqls_2} and ansatz variations such as the Logical Ansatz, being a Classical Combination of various Ans\"{a}tze \cite{logical_ansatz, vqls_2} and the Evolutionary Ansatz \cite{evqls}, an evolutionary approach for Ansatz construction. The evolutionary ansatz was initially propos\-ed for use in the VQE and has been applied here for the VQLS variation.  Another non-variational approach to solving the quantum linear systems prob\-lem is the Classical Combination of Quantum States (CQS) method \cite{vqls_2}, of which the Logical Ansatz approach outlined in this paper is an adaption.

The following approaches were implemented and will be discussed in this paper:
\begin{center}
\begin{tabular}{ll} 
 Variational Quantum Linear Solver & (VQLS)  \\ 
 Adiabatic Ansatz VQLS & (AAVQLS) \\ 
 Evolutionary Ansatz VQLS & (EAVQLS)\\ 
 Classical Combination of Quantum State & (CQS) \\
 Logical Ansatz VQLS & (LAVQLS) \\
\end{tabular}
\end{center}

This paper makes several contributions to the literature on the VQLS. Firstly, we present the first application of the Evolutionary Ansatz to the VQLS. The Evolutionary Ansatz has previously been applied to the Variational Quantum Eigensolver \cite{evqls}. 
Secondly, the implementation proposed for solving systems of linear equations using the AAVQLS is also new to this work.  
Also, the first proof of principle demonstration of the CQS method on a real quantum device is conducted.
Lastly, the first known implementation of the Logical Ansatz VQLS is given, with proposed training methods that are new to this work.
All implementations of these approaches may be found in the Github repository \cite{git}.

This paper begins with a description of the above near-term approaches for the Quantum Linear Systems Problem. Some experiments designed for an evaluation of these approaches are outlined in the section following that. We then present and discuss the results of the experiments.

\section{Near-Term Algorithms}
\label{near_term_approaches}
The inputs to the all the near-term algorithms below are the matrix $\mathbf{A}$, and vector $\vec{b}$.
$\mathbf{A}$ is given in a slightly different form here than in the QLSP. Here $\mathbf{A}$ is given by $m$ unitary matrices $A_{i}$, implemented as unitary gates, such that $\mathbf{A} = \sum_{i=1}^{m}c_{i}A_{i}, c_{i} \in \mathbb{C}$ (any hermitian matrix in a finite-dimensional space can be written as a linear combination of unitary matrices). Also given is a unitary gate $\textit{U}$, such that $\textit{U}\ket{0} = \ket{b}$.  VQLS cost functions often require either or both $A_i$ and $U$ to be given as controlled gates, which is assumed possible.

\subsection{Variational Quantum Linear Solver}
\label{standard_vqls}

The standard VHQCA approach for the quantum linear systems problem is the Variational Quantum Linear Solver, itself being a basic application of the VQE. The VQLS simply involves the selection of a suitable ansatz, cost function, and classical optimizer. The algorithm runs in a simple feedback-loop, whereby the classical optimizer finds the optimal parameters for the ansatz circuit, by iteratively evaluating the cost function on the quantum device, and updating the parameters until a minimum cost value is achieved. The quantum device is used to evaluate the cost function, because it is much more efficient than any known method on a classical device for this step \cite{vqls_1}.

Let the ansatz be denoted by $V(\mathbf{\alpha})$, and let the optimal ansatz parameters be denoted by $\alpha^{*}$. Then once the VQLS algorithm terminates, $V(\mathbf{\alpha}^{*})\ket{0} = \ket{x'}$, where  $\lvert\lvert\ket{x}-\ket{x'}\rvert\rvert_{2}  \leq \epsilon$, where $\epsilon$ is the error-bound of the approximate solution, and $\ket{x}$ is the exact solution as described by equation (\ref{equation_1}).

\begin{figure}[h]
\centering
\includegraphics[width=\columnwidth]{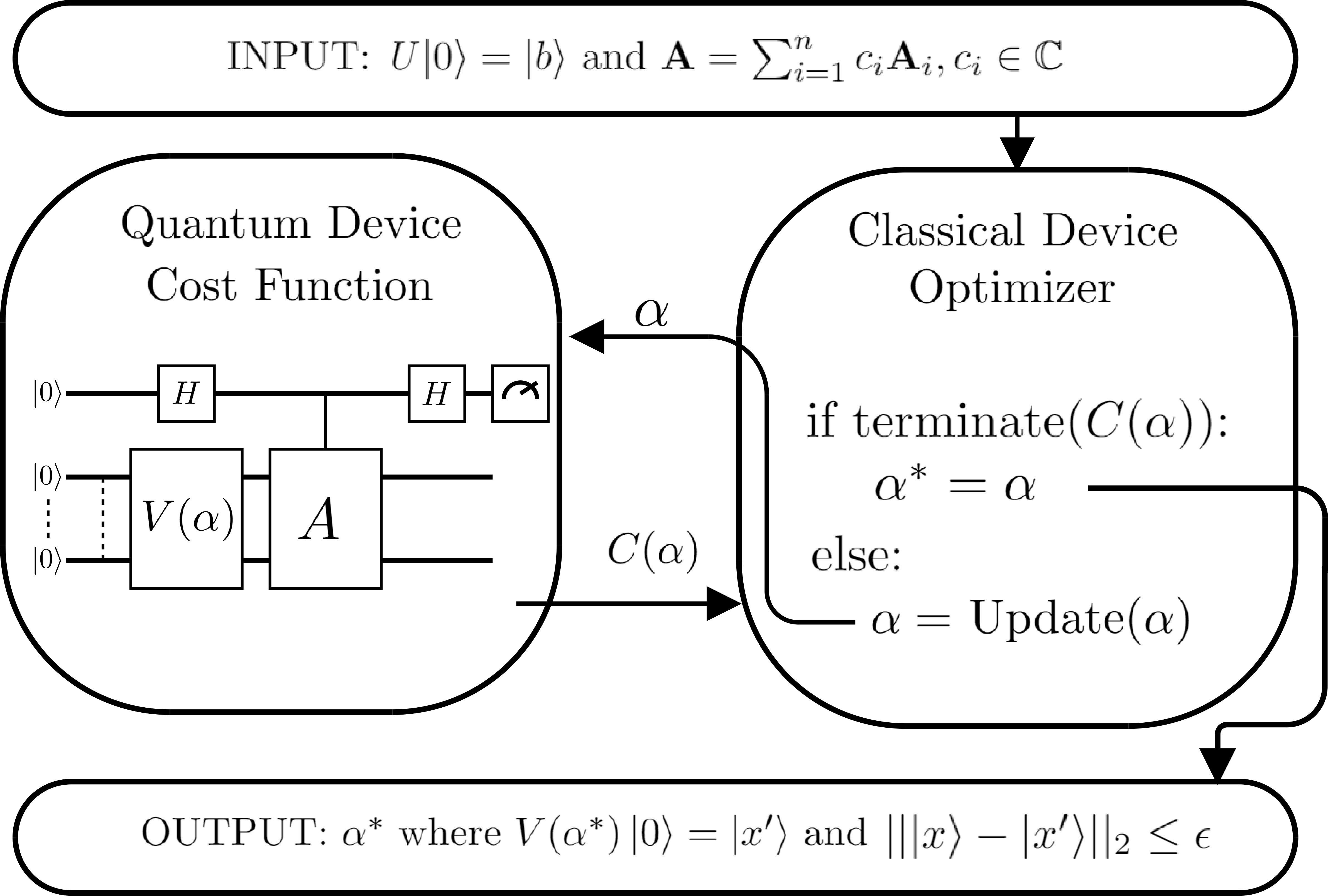}
\caption{VQLS Schematic: The algorithm runs in a simple feedback loop, whereby a classical optimizer finds the optimal parameters for the ansatz circuit, by iteratively evaluating the cost function on the quantum device and updating the parameters, until a minimum cost value is achieved. $V(\mathbf{\alpha})$ denotes the parameterized ansatz, $\mathbf{\alpha}^{*}$ denotes the optimal parameters. At termination,  $V(\mathbf{\alpha}^{*})\ket{0} = \ket{x'}$, where $\lvert\lvert\ket{x}-\ket{x'}\rvert\rvert_{2}  \leq \epsilon$, with $\ket{x}$ being the exact solution to the QLSP, and $\epsilon$ being the error-bound of the approximate solution.}
\end{figure}

\subsubsection{VQLS Ansatz}
\label{standard_vqls_ansatz}
Broadly speaking there are two types of ans\"{a}tze; hardware-efficient (agnostic) ans\"{a}tze and problem-specific ans\"{a}tze. 

Hardware-efficient ans\"{a}tze are designed without taking into account the specific problem being solved, that is matrix $\mathbf{A}$ and $\ket{b}$, but rather only the topology (backend connectivity of the qubits) and available gates of a specific quantum computer. A hardware-efficient ansatz can be denoted by a sequence of $n$ param\-eterized quantum gates as, 
\begin{center}
\begin{gather}
     V_{\mathrm{Agnostic}}(\alpha) = \gamma_{1}(\alpha_{1})\gamma_{2}(\alpha_{2}) \cdots \gamma_{n}(\alpha_{n}),
\end{gather}
\end{center}
where $\gamma_{i}$ denotes a specific parameterized gate in the quantum circuit, and  $\alpha_{i}$ denotes that parameters value.

These ans\"{a}tze can be constructed to be more resistant to noise on any specific available quantum device, but they may fall short finding a solution $\ket{x'}$, as any particular hardware-efficient ans\"{a}tze is not guaranteed to span the region of the Hilbert space containing any good approximation of the solution $\ket{x'}$. Therefore a hardware-efficient ansatz effectively trades potential relevance to the specific problem, for increased noise resistance.

Problem-specific ans\"{a}tze on the other hand do not take into account the specific quantum device being used, and rather try to exploit the knowledge of the problem available. The Quantum Alternating Operator Ansatz (QAOA) \cite{vqls_1} is one such proposed problem-specific ansatz, using two hamiltonians, known as the \emph{driver} and the \emph{mixer}, denoted by $H_D$ and $H_M$ respectively, constructed from specific knowledge of the problem, namely $\mathbf{A}$ and $\vec{b}$. This problem-specific ansatz can be denoted by a repeating sequence of \emph{driver} and \emph{mixer} hamiltonian simulations, each being applied for a variable amount of time. These time parameters $\mathbf{\alpha}$ are the trainable aspect of this problem specific ansatz, which are optimized by some classical device. The QAOA can be denoted as,
\begin{center}
\begin{gather}
     V_{QAOA}(\alpha) = e^{-iH_{M}\alpha_{2p}} e^{-iH_{D}\alpha_{2p-1}} \dots e^{-iH_{M}\alpha_{2}} e^{-iH_{D}\alpha_{1}}.
\end{gather}
\end{center}

The requirement of Hamiltonian simulation from the QAOA makes these ans\"{a}tze far less near-term, therefore these ans\"{a}tze are not considered further in this paper. More information on the specific construction of the QAOA, including operators $H_D$ and $H_M$ is given in \cite{vqls_1}.

\subsubsection{VQLS Cost Functions}
\label{standard_vqls_ansatz}

\begin{figure}[h]
\centering
\includegraphics[width=\columnwidth]{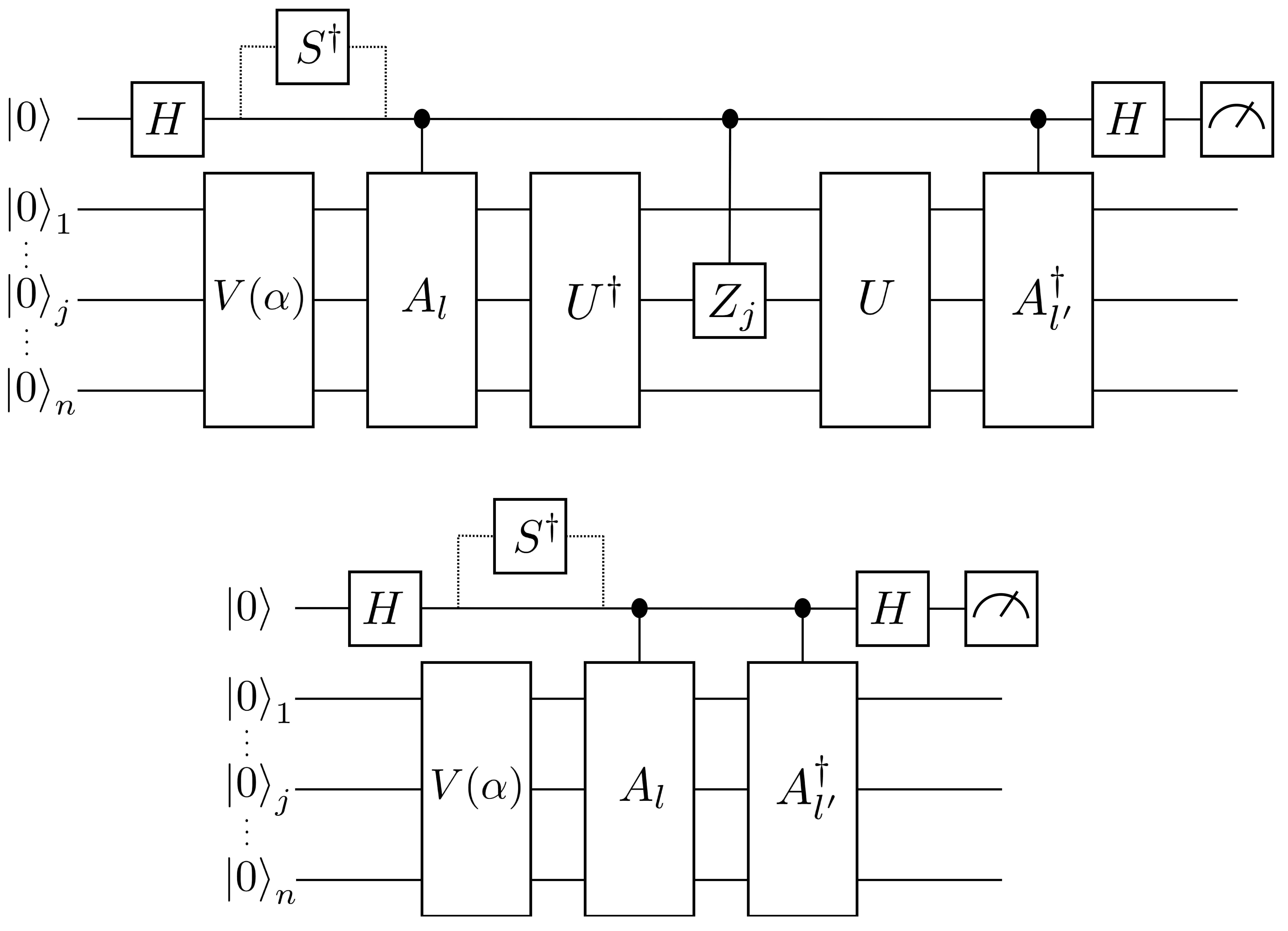}  
\caption{Hadamard Test Circuits for cost function $C_{L}$ Eq. (\ref{cl}).
The top circuit is employed when calculating the value of the numerator $\expval{H_L}{x}$, while the bottom circuit is employed when calculating the value of the denominator, $\bra{\psi}\ket{\psi}$. The $S^\dagger$ gate, is included when calculating imaginary-valued parts of the cost function, and excluded when calculating the real-valued parts.}
\label{had_test}
\end{figure}

The cost function Hamiltonian is where the application of the VQE algorithm to solving systems of linear equations is implemented. Various different cost functions have been proposed for the VQLS \cite{vqls_1, vqls_2}. %For simplicity, denote the state created by the ansatz when applied to $\ket{0}$, namely $V(\alpha)\ket{0}$, as $\ket{x}$, and let $\ket{\psi} = A\ket{x}$.%
For simplicity, denote the state $V(\alpha)\ket{0}$, as $\ket{x}$, and let $\ket{\psi} = A\ket{x}$. Ref. \cite{vqls_1} proposes a cost function based on the overlap between the projector $\ketbra{\psi}{\psi}$ and the subspace orthogonal to $\ket{b}$. This cost function also normalizes the expectation value of the Hamilton\-ian to improve performance. The cost function is given by,
\begin{center}
\begin{gather}
     C_G = \dfrac{\expval{H_G}{x}}{\bra{\psi}\ket{\psi}},
\end{gather}
\end{center}
where the Hamiltonian $H_G$ is given by,
\begin{center}
\begin{gather}
     H_G = A^{\dagger}(\mathbbm{1} - \ketbra{b}{b}))A.
\end{gather}
\end{center}
This  cost function can have gradients that vanish exponentially with the number of qubits. To improve on this shortfall, the cost function $C_{L}$ is proposed by replacing $H_{G}$ with a local version of the Hamiltonian, $ H_{L} $, improving the trainability of the ansatz, given by,
\begin{center}
\begin{gather}
     H_{L} = A^{\dagger}U(\mathds{1}-\frac{1}{n}\sum^{n}_{j=1}\ketbra{0_{j}}{0_{j}}\otimes\mathds{1}_{\Bar{j}})U^{\dagger}A,
\end{gather}
\end{center}

where $\mathds{1}_{\Bar{j}}$ denotes identity on all qubits except qubit $j$.
%[] proposes cost functions that aim to minimize the value of $\lvert\lvert \mathbf{A}\ket{x} - \ket{b}\rvert\rvert_2$. %\textit{\\elborate here on above paper method and spacing}
The cost function $C_{L}$ can be computed using the Hadamard Test as shown in Fig. \ref{had_test}. $C_{L}$ has been shown to be equivalent cost function to $C_{G}$, however having improved performance \cite{vqls_1}, and is explicitly given by, 

\begin{center}
\begin{gather}
\label{cl}
     C_L = \dfrac{\expval{H_L}{x}}{\bra{\psi}\ket{\psi}}.
\end{gather}
\end{center}

\subsubsection{Classical Optimizers}
\label{standard_vqls_classical_optimizer}
The VQLS admits the use of either gradient-free or gradient-based optimizers. For gradient-based optimizers, gradient values can be found analytically \cite{vqls_1, gradient}, or estimated through finite differences. The classical optimi\-zer chosen has a large impact on how well the optimization process deals with the noise inherent in NISQ devices. Some classical optimizers handle noise better than others \cite{paper_1}, making classical optimizer selection important. % The hyper-parameter settings of the classical optimizers are also of notable importance. 

\subsection{Adiabatic Assisted VQLS}
The Adiabatic Assisted Variational Quantum Linear Solver (AAVQLS) \cite{vqls_2}, simply augments the standard VQLS approach, by proposing a variation in the Hamil\-tonian over time, inspired by adiabatic quantum computing methods \cite{adiabatic}, in an attempt to allow the ansatz state to always be close to the ground state of the Hamiltonian. Let $H_0$ be a Hamiltonian with a known ground state, and let $H_1$ be the Hamiltonian whose ground state corresponds to the solution of the linear system in question. Let the Hamiltonian of the AAVQLS be given by $(1-s)H_0 + sH_1$, where $s$ is a discrete parameter, varying from $s = 0$, to $s = 1$, in $T$ discrete intervals, during the optimization process. 

This approach is the same as the VQLS with respect to the cost function and classical optimizer, however the ansatz must be chosen such that it can be easily initialized in the ground state of $H_0$ at the start of the algorithm. The only added procedure to the AAVQLS from the VQLS occurs during the training phase, where the parameter $s$ is varied in $T$ discrete intervals from $s = 0$ to $s = 1$, thereby varying the hamiltonian from $H_0$ to $H_1$.

Proposed here is one way in which the AAVQLS can be implemented as an adaption of the VQLS. Firstly the linear system is reformulated as,
\begin{center}
\begin{gather}
[(1-\overline{s})\mathbbm{1} + \overline{s}\mathbf{A}]\vec{x} = \vec{b},
\end{gather}
\end{center}
where $\overline{s}$ can be varied from $0$ to $1$, with $\vec{x} = \vec{b}$, when $\overline{s} = 0$, and $[(1-\overline{s})\mathbbm{1} + \overline{s}\mathbf{A}]\vec{x} = \vec{b}$, equivalent to $\mathbf{A}\vec{x} = \vec{b}$, when $\overline{s} = 1$.

Then for a suitable ansatz $V(\mathbf{\alpha}) = \gamma_{1}(\alpha_{1})\gamma_{2}(\alpha_{2}) \cdots \gamma_{n}(\alpha_{n})$ where $\mathbf{\alpha}$ can be initialized such that $V(\mathbf{\alpha}) = \mathbbm{1}$, append to it the unitary $\textit{U}$ (for creating state $\ket{b}$) to create ansatz $V_{AAVQLS}(\mathbf{\alpha})$ as below,
\begin{center}
\begin{gather}
V_{AAVQLS}(\mathbf{\alpha}) = UV(\mathbf{\alpha}).
\end{gather}
\end{center}
Here $V_{AAVQLS}\ket{0}$ is indeed be equal to $\ket{b}$ when $\mathbf{\alpha}$ is initialized appropriately (such that $V(\mathbf{\alpha}) = \mathbbm{1}$). 

%Using any implementation of the standard VQLS, by varying $\overline{s}$ discretely during %optimization from $0$ to $1$, the solution to the system of linear equations can be found.

The AAVQLS algorithm then proceeds as follows:
\begin{enumerate}
    \item Let $\overline{s} = 0$ and let ansatz parameters $\mathbf{\alpha} = \mathbf{\alpha*}$, such that $V(\mathbf{\alpha*})$ is equal to $\mathbbm{1}$. Then $V_{\mathrm{AAVQLS}}(\mathbf{\alpha})$ gives the solution to (9). Let $\alpha_{\mathrm{previous}}=\alpha$.
    
    \item Increment $\overline{s}$  by $\frac{1}{T}$.
    
    \item Using the VQLS approach, with initial ansatz parameters set to $\alpha_{\mathrm{previous}}$, find the new optimal parameters $\alpha_{\mathrm{current}}$ with ansatz $V_{\mathrm{AAVQLS}}$.
    
    \item Let $\alpha_{\mathrm{previous}} = \alpha_{\mathrm{current}}$ and if $\overline{s}\neq1$ return to step 2; else $\alpha_{\mathrm{final}}=\alpha_{\mathrm{current}}$.
    
    \item $V_{\mathrm{AAVQLS}}(\mathbf{\alpha}_{\mathrm{final}})\ket{0} = \ket{x'}$, where $\lvert\lvert\ket{x}-\ket{x'}\rvert\rvert  \leq \epsilon$, where  $\ket{x}$ is the exact solution to the QLSP and $\epsilon$ is the error-bound of the approximate solution. 
\end{enumerate}

\subsection{Evolutionary Ansatz VQLS}
The Evolutionary Ansatz \cite{evqls} utilises a genetic algorithm to construct the paramet\-erized quantum circuit, while concurrently optimising its parameters. This approach adapts the ansatz structure to both the specified problem and the backend configuration of the quantum device available, so it can be thought of as construc\-ting an ansatz that is both hardware-efficient and problem-specific. This specialized ansatz would be highly noise resistant (being shallower and requiring fewer 2-qubit gates) while remaining applicable to the specific problem being solved. This approach still requires a VQLS cost function and classical minimizer, as standard in with VQLS. 

The main outline of the Evolutionary Ansatz algorithm is presented here. For a full explanation please see the original paper\cite{evqls}. Some specific details about the evolutionary ansatz implementation in this discussion differ from the original paper.

\begin{figure}[h]
\centering
\includegraphics[width=\columnwidth]{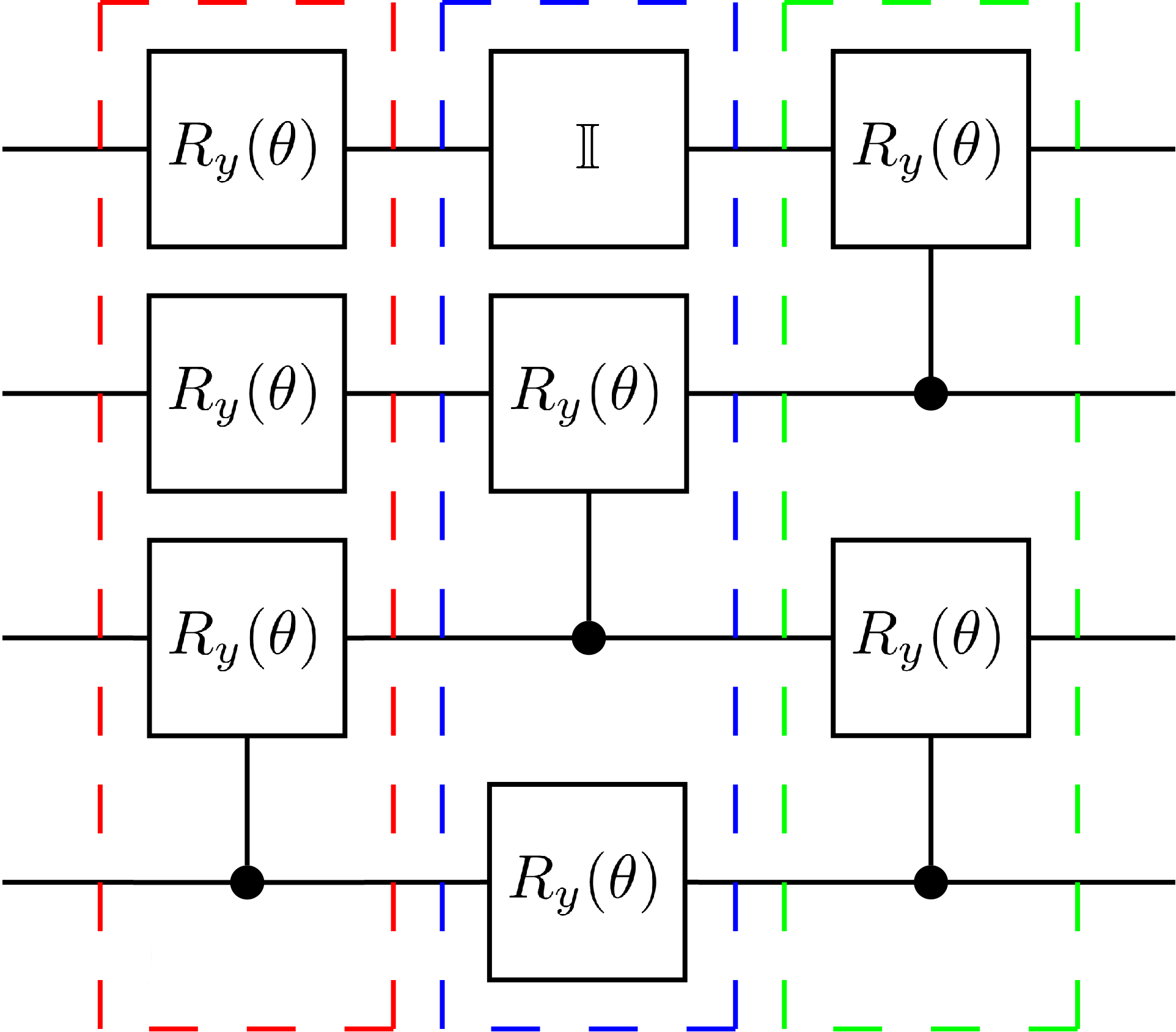}
\caption{4 Qubit Evolutionary Ansatz Genome Schematic: Three different genes (ansatz rows) are each outlined in red, blue and green in the genome above. The qubits are initialized as $H^{\otimes{n}}\ket{0}^{\otimes{n}}$, allowing the first gene in the genome to contain controlled gates that aren't redundant.
}
\label{eansatz}
\end{figure}

Evolutionary Algorithms are based on the principle of natural selection. The Evolutionary Ansatz VQLS (EAVQLS) mimics this principle to adapt the ansatz choice by evolving a set of candidate ansatze, known as the population, through random mutations (the EAVQLS only mimics asexual reproduction, there are no crossovers between candidate solutions in the population). The candidate ansatze being evolved are known as genomes. A genome consists of a list of genes, and for the EAVQLS is given as follows. If $V_{i}(\alpha)$ is any potential ansatz in the population, $V_{i}(\alpha)$ is expressed as a genome $g_{i}$ that can be written as, 
\begin{center}
$g_{i} = [\gamma_{1}(\alpha_{1}), \gamma_{2}(\alpha_{2}), \cdots, \gamma_{m}(\alpha_{m})]$,
\end{center}
where each $\gamma_{i}$ is a gene. Each gene $\gamma_{i}$ is a layer of gates such that each qubit of the ansatz is assigned a gate. This set of gates is chosen from a gate set,
\begin{center}
$\mathbb{G} = \{\mathbb{I}_{2}, U_3, \wedge_{1}U_3\},$
\end{center}
where $\mathbb{I}_{2}$ is the single qubit identity gate, $U_3$ is the universal single qubit rotation, and $\wedge_{1}U_3$ is the controlled version of $U_3$. Other gate sets may be used, for example, if the VQLS problem only deals with real valued $\mathbf{A}$ and $\vec{b}$, an appropriate gate set is given by,
\begin{center}
$\mathbb{G_R} = \{\mathbb{I}_{2}, R_{y}, \wedge_{1}R_{y}\}.$
\end{center}
An example of an evolutionary ansatz genome is shown in Fig. \ref{eansatz}.

Initially, the evolutionary algorithm begins with a population consisting of random genomes, each only consisting of a single gene. The qubits in all ansatz circuits are initialized with the gate $H^{\otimes{n}}$, allowing the first gene in the genome to contain controlled gates that aren't redundant when the ansatz corresponding to the genome is applied to the state $\ket{0}^{\otimes{n}}$. These genomes are evolved asexually, through the use of 3 genetic operators, \emph{topological search}, \emph{parameter search} and \emph{removal}.

The \emph{topological search} operator, $\tau$, explores the space of possible ans\"atze, by adding a new random gene to the genome, which is equivalent to adding a new random layer of gates to the ansatz represented by the genome. The new gene added to the genome is initialized as identity, to ensure that the genome's fitness may only improve, or at worst, remain the same. The new gene added to the genome also takes into account the gates of the previous gene in the genome, eliminating potential redundant gates (two of the same gate on the same qubit/s in a row) being added to the genome with the new gene. The identity gate, $\mathbb{I}_{2}$, is only used whenever adding a different gate would cause some redundancy. The operation performed by $\tau$ is given by, 
\begin{center}
    $\tau:[\gamma_{1}(\alpha_{1}), \cdots, \gamma_{m}(\alpha_{m})] \xrightarrow{} [\gamma_{1}(\alpha_{1}), \cdots, \gamma_{m}(\alpha_{m}), \gamma_{m+1}(\alpha_{m+1})]$.
\end{center}

The \emph{removal} operator, $\rho$, acting on a genome, removes some number of genes from the genome, starting from the end of the gene list. The operation performed by $\rho$ is given by,  

\begin{center}
    $\rho:[\gamma_{1}(\alpha_{1}), \cdots, \gamma_{p}(\alpha_{p}), \cdots, \gamma_{m}(\alpha_{m})] \xrightarrow{} [\gamma_{1}(\alpha_{1}), \gamma_{2}(\alpha_{2}), \cdots, \gamma_{p}(\alpha_{p})]$
\end{center}
where $p \in \{1, 2, \cdots, m\}$.

The \emph{parameter search} runs an optimization subroutine, $\mathbb{O}$, on each individ\-ual gene in the genome, in a random order. The optimization subroutine $\mathbb{O}$ is implemented for a genome $g_i$ and a gene $y_i(\alpha_i)$, by using the ansatz $V(\alpha)$ represented by genome $g_i$ in the standard VQLS optimization routine, while only optimizing the specific parameters $\alpha_i$ of the gene $y_i(\alpha_i)$. This optimization is done per gene so that \emph{removal} operator does not affect the training of the rest of the ansatz. 

The fitness $f_{i}$, of genome $g_{i}$ is calculated using the value of the cost achieved by the ansatz represented by the genome, as well as the depth of the ansatz (number of genes) and the number of 2-qubit gates. The fitness value is given by,
\begin{center}
    $f_{i} = \mathbb{C}(g_{i}) + \alpha\cdot\lvert g_{i}\rvert + \beta\cdot \lvert \wedge(g_{i})\rvert$,
\end{center}
where $\alpha$ and $\beta$ are weighted variables that can be assigned, $\mathbb{C}(g_{i})$ is the cost value of the ansatz of represented by $g_{i}$, $\lvert g_{i}\rvert$ is the number of genes in $g_{i}$, and $\lvert \wedge(g_{i})\rvert$ is the number of 2-qubit gates in $g_{i}$. This genome fitness value is then averaged amongst genomes of the same species and is called the species-adjusted fitness. Species are defined by a genetic distance measure, given by the average distance of a common ancestor between two separate genomes. Two genomes with an average distance of a common ancestor less than some given value may be assigned to the same species.

The EAVQLS Algorithm runs as follows:
\begin{enumerate}
    \item Generate a population $P$ of $n$ empty genomes $g_i$, and apply $\tau(g_i)$ to each genome. 
    \item Apply the optimization subroutine $\mathbb{O}$ to the last gene in each genome for all genomes in $P$.
    \item  Group the genomes in $P$ by species, then calculate the fitness and then species-adjusted fitness of each genome $g_i$ in $P$.
    \item Randomly select $n$ parent genomes with replacement from $P$, inversely proportional to their fitness values, forming the next generation $P'$. 
    \item For all of the genomes $g_i$ in $P'$, apply the three genetic operators to $g_i$ with some predefined probabilities.
    \item If a termination condition is met, return the fittest genome in $P'$, else let $P = P'$ and return to step 2. 
\end{enumerate}

\subsection{Classical Combination of Quantum States (CQS)}
The Classical Combination of Quantum States (CQS) approach detailed in \cite{vqls_2}, is the most unique near-term approach presented in this paper. The CQS approach is not a variational algorithm, meaning there is no classical optimisat\-ion of ansatz parameters. The CQS approach solves the linear system by finding the solution as a linear combination of quantum states.

Given a set of $n$ states $S=\{\ket{s_1}, \cdots, \ket{s_n}\}$, the CQS method aims to find a linear combination $x'$ approximating the solution $x$ to the linear systems problem $A\vec{x}=\vec{b}$ where,
\begin{center}
\begin{gather}
x' = \sum_{i=1}^n \alpha_{i}\ket{s_{i}}, \alpha \in \mathbb{C}.
\end{gather}
\end{center}

Note that, differing from the above mentioned approaches, the solution $x'$ is never actually created on the quantum device. It is not necessarily normalized, and is proportional to the solution to the same problem solved using the other variational methods.

The CQS Algorithm runs as follows. Starting with $m = 1$, and the set $S = \{\ket{s_1}\}$ where $\ket{s_1}$ is a state that can be prepared by some efficient quantum circuit.
\begin{enumerate}
    \item Solve the for the optimal values of $\mathbf{\alpha^*}$ such that $x' = \Sigma_{i=1}^m \alpha^*_i \ket{s_i}$, where $x'$ is the an approximation of $x$, given the set $S$.
    \item Using the value of $\mathbf{\alpha^*}$, find the next circuit generating the state $\ket{s_{m+1}}$ and add $\ket{s_{m+1}}$ to $S$.
    \item Set $m = m + 1$ and repeat from step 1 until $x'$ is a sufficiently good solution. 
\end{enumerate}

Given a set of $n$ efficiently prepared states, $S=\{\ket{s_1}, \cdots, \ket{s_n}\}$, containing a close approximation of the solution $x$ as a linear combination, the linear coefficients, $c_{i}$, can be found efficiently by a hybrid quantum-classical procedure outlined below.

The standard regression function used to solve a linear system is given by,
\begin{center}
\begin{gather}
L_R(x) \coloneqq \lvert\lvert Ax - \ket{b} \rvert\rvert_{2}^{2} = x^{\dagger}A^{\dagger}Ax - 2\text{Re}[\bra{b}Ax] + 1.
\label{L_r}
\end{gather}
\end{center}
Given $x = \sum_{i = 1}^{m}\alpha_i\ket{s_i}$, let $V = (v_1, \dots, v_m)$ where $v_i = A\ket{s_i}$. Now $Ax = \sum_{i=1}^{m}\alpha_i A\ket{s_i} = V\alpha$. Eq. (\ref{L_r}) can be rewritten as,
\begin{center}
\begin{gather}
\lvert\lvert V\alpha - \ket{b} \rvert\rvert_{2}^{2} = \alpha^{\dagger}V^{
\dagger}V\alpha - 2\text{Re}\{q^{\dagger}\alpha\} + 1.
\end{gather}
\end{center}
where $q_i = \bra{i}V^{\dagger}\ket{b} = \bra{s_i}A^{\dagger}\ket{b}$. A simple regression problem for $\alpha$ can be obtained with the kernel matrix $V^{\dagger}V$, where $(V^{\dagger}V)_{ij} = \bra{s_i}A^{\dagger}A\ket{s_j}$. This quadratic optimization problem on complex $\alpha$ can be rewritten as a real optimiza\-tion problem,
\begin{center}
\begin{gather}
\text{min}_z (z^{\dagger}Qz - 2r^Tz + 1),
\label{opt}
\end{gather}
\end{center}
 where $z = (\text{Re}[\alpha], \text{Im}[\alpha])$, and $Q$ and $r$ are given by,
\begin{equation}
Q = 
\begin{pmatrix}
\text{Re}[V^{\dagger}V] & \text{Im}[V^{\dagger}V] \\
\text{Im}[V^{\dagger}V] & \text{Re}[V^{\dagger}V]
\end{pmatrix},
\end{equation}

\begin{equation}
r =
\begin{pmatrix}
\text{Re}[q] & \text{Im}[q]
\end{pmatrix}.
\end{equation}

Here Eq. (\ref{opt}) can be solved using standard convex quadratic programming methods.

\begin{figure}[h]
\centering
\includegraphics[width=\columnwidth]{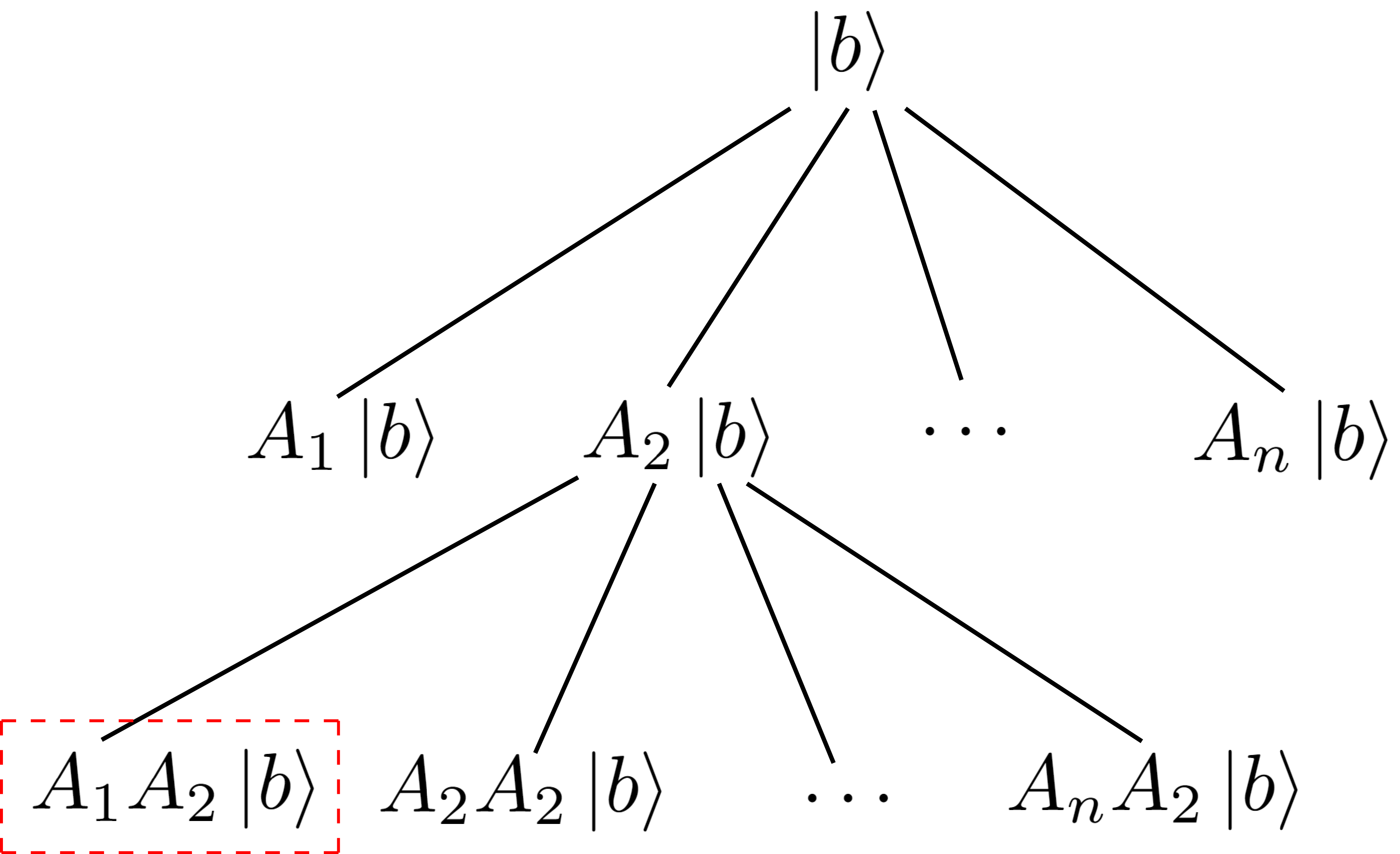}
\caption{Ansatz Tree Diagram: An example expansion of the Ansatz Tree. The highlighted node in the tree represents the unitary created by applying the unitaries $A_2$ and then $A_1$ to the state $\ket{b}$.}
\label{atree}
\end{figure}

The Ansatz Tree (Fig. \ref{atree}) is a proposed structure used to obtain a good set of unitaries. As previously specified, the matrix $A$ is given as a linear combination, $A = \sum_{i=1}^{n}c_{i}A_{i}, c_{i} \in \mathbb{C}$. The unitaries $A_{i}$ are used in the construction of the ansatz tree as follows.

Each node in the ansatz tree represents a single state, $\ket{s_i}$, composed from the unitaries $\mathbf{A}_{i}$ and $\ket{b}$, which can be added to the ansatz set used to find the linear combination on a classical device. 
The ansatz tree can be expanded breadth-first, or searched via some heuristic.

A heuristic approach to searching the ansatz tree also proposed in \cite{vqls_2}, aims to expand the ansatz tree node by node. Let the current set of expanded nodes in the tree be given by the set $S$, and the current set of all potential child nodes of the nodes in $S$ be the set $\mathit{C}(S)$. Let the current approximate solution for the set of expanded nodes be given by $\mathbf{\alpha^*}$. The ansatz tree is further explored by expanding the child node, $\ket{\psi} \in \mathit{C}(S)$, that has the greatest overlap with the current gradient, maximizing the function given by,
\begin{center}
\begin{gather}
    \bra{\psi}\nabla L_{R}(x^{s}) = 2\sum_{\ket{\psi_i}\in S}\alpha_i^* \bra{\psi} A^{2}\ket{\psi_i} - 2\bra{\psi} A\ket{b}.
\end{gather}
\end{center}

\subsection{Logical Ansatz VQLS}
\label{la_section}
The logical ansatz approach is simply an adaption of the CQS method above, allowing for parameterized unitaries to be employed. This approach is then similar to the Standard VQLS approach, except it proposes that instead of a single ansatz circuit, a linear combination of multiple ansatz circuits be used. This greatly lowers the depth of any one of the multiple ansatz circuits. The Logical Ansatz is implemented here as suggested in \cite{vqls_2}, by implementing the CQS method with a selected set of $n$ parameterized ansatze making the states $S = \{\ket{s_1(\mathbf{\theta}_1)}, \cdots, \ket{s_n(\mathbf{\theta}_n)}\}$, and repeating the optimization process outlined for the CQS with a classical minimizer optimizing the parameters $\mathbf{\theta}_i$ of the ansatze creating the states $\ket{s_i(\mathbf{\theta}_i)}$. This method avoids the ansatz tree expansion process for selecting unitaries, by instead optimizing a set of pre-selected parameterized unitaries. The solution found is expressed by, 
\begin{center}
\begin{gather}
x' = \sum_{i=1}^n \alpha_{i}\ket{s_{i}(\mathbf{\theta}_i)}, \alpha_i \in \mathbb{C}.
\end{gather}
\end{center}
This logical ansatz implementation and training differs from that in \cite{logical_ansatz}.

The Logical Ansatz Linear Solver Algorithm runs much like the CQS method. 
\begin{enumerate}
    \item Select $n$ parameterized ansatze each corresponding to some state $\ket{s_i(\theta_i)}$, forming set $S$. 
    \item Solve the for the optimal value of $\mathbf{\alpha^*}$ such that $x' = \Sigma_{i=1}^m \alpha^*_i\ket{s_i(\theta_i)}$, where $x'$ is the closest approximation of $x$, given the set $S$. Proceed either to 3 or 4 for method 1 or 2 respectively.
    \item Method 1: For $r$ training rounds, select each of the states $\ket{s_i(\theta_i)}$ at random and optimize their parameters $\theta_i$ with some classical optimizer, only solving the new $\mathbf{\alpha^*}$ value after the parameter optimization of each ansatz.
    \item Method 2: Treating the entire state $x' = \Sigma_{i=1}^m \alpha^*_i\ket{s_i(\theta_i)}$ as a single logical ansatz, optimize all parameters at once with a classical optimizer, solving for the new $\mathbf{\alpha^*}$ value with each change of parameter during the optimization process. 
\end{enumerate}

\section{Experimentation and Evaluation}
\label{results}
Tests of all above described methods follow below. The linear systems to be solved are given as a matrix $\mathbf{A}$, where $\mathbf{A} = \Sigma_l c_l A_l$, for $l$ unitary gates, and a state $\ket{b}$, given as a unitary $U$, prepared by some efficient quantum circuit, such that $U\ket{0} = \ket{b}$, corresponding to some $\vec{b}$, as described in the formulation of the near-term Quantum Linear Systems Problem. 

In all problem instances detailed below, the state $\ket{b} = H^{\otimes n}\ket{0}^{n}$, where $n$ is the relevant number of qubits for the problem. These problems are also only real-valued linear systems, however these approaches are not limited to real-values. The number of shots for the Qiskit's Qasm simulator is set to 10000 (except for the CQS approach on the real device). 

\subsection{Variational Quantum Linear Solver}
\label{vqls_results}

Three problem instances of differing sizes were selected
to investigate the perform\-ance of the standard VQLS. Two different classical optimizers (gradient-based vs gradient-free), and three levels of noise were tested in order to further investigate their role in the performance of the VQLS. The two chosen optimizers were the gradient-based Broyden–Fletcher–Goldfarb–Shanno algo\-rithm (BFGS) \cite{bfgs} (using an analytic gradient function computed on the quantum hardware) and the gradient-free Simultaneous Perturbation Stochastic Approxi\-mation algorithm (SPSA) \cite{spsa}, based on performance in \cite{paper_1}. The three noise levels were chosen such to demonstrate the difference between zero noise, shot-noise only and realistic NISQ device noise, given by Qiskit's Statevector simulator, Qasm simul\-ator and Qasm simulator with a realistic noise sample respectively. The noise sample is taken from the IBM-Vigo quantum device.

The three problem instances, using 3, 4 and 5 qubits respectively, are given by,
\begin{center}
$\mathbf{A}_{1} = H_{1} + 0.25\cdot Z_{2} + 0.15\cdot H_{3}$,

$\mathbf{A}_{2} = Z_{1} + 0.25\cdot Z_{2} + 0.5\cdot Z_{4},$

$\mathbf{A}_{3} = H_{1} + 0.25\cdot Z_{3} + 0.5\cdot H_{5},$
\end{center}
where $Z_{i}$, ($i=1,2,3$) indicates the matrix formed by the tensor product, with Pauli gate $Z$ applied to qubit $i$ and the identity gate applied to the remaining qubits. Notation is similarly defined with Hadamard gate $H$ and Pauli gate $X$. $\mathds{1}$ indicates an $N \times N$ identity matrix. 

The local cost function (detailed above) was selected for all the VQLS runs. 100 runs of the VQLS algorithm were conducted for each problem instance, noise-level and classical optimizer. Furthermore, the same 100 random initial ansatz parameter values were used for all runs in the same problem instance across all noise levels and classical optimizers. This was done to make the results obtained for each problem instance comparable.

In order to ensure even resource distribution between the gradient-based BFGS and the gradient-free SPSA classical optimizers, the optimizers were only allowed a limited number of cost function evaluations. The resource cost of a gradient call can be given in terms of cost function calls, being exactly 2 cost function calls per ansatz parameter, and so this comparison can be done. The 3, 4 and 5 qubit problem instances were limited to 1000, 1500 and 2000 cost function evaluations respect\-ively. 

\begin{figure}[h]
\centering
\includegraphics[width=\columnwidth]{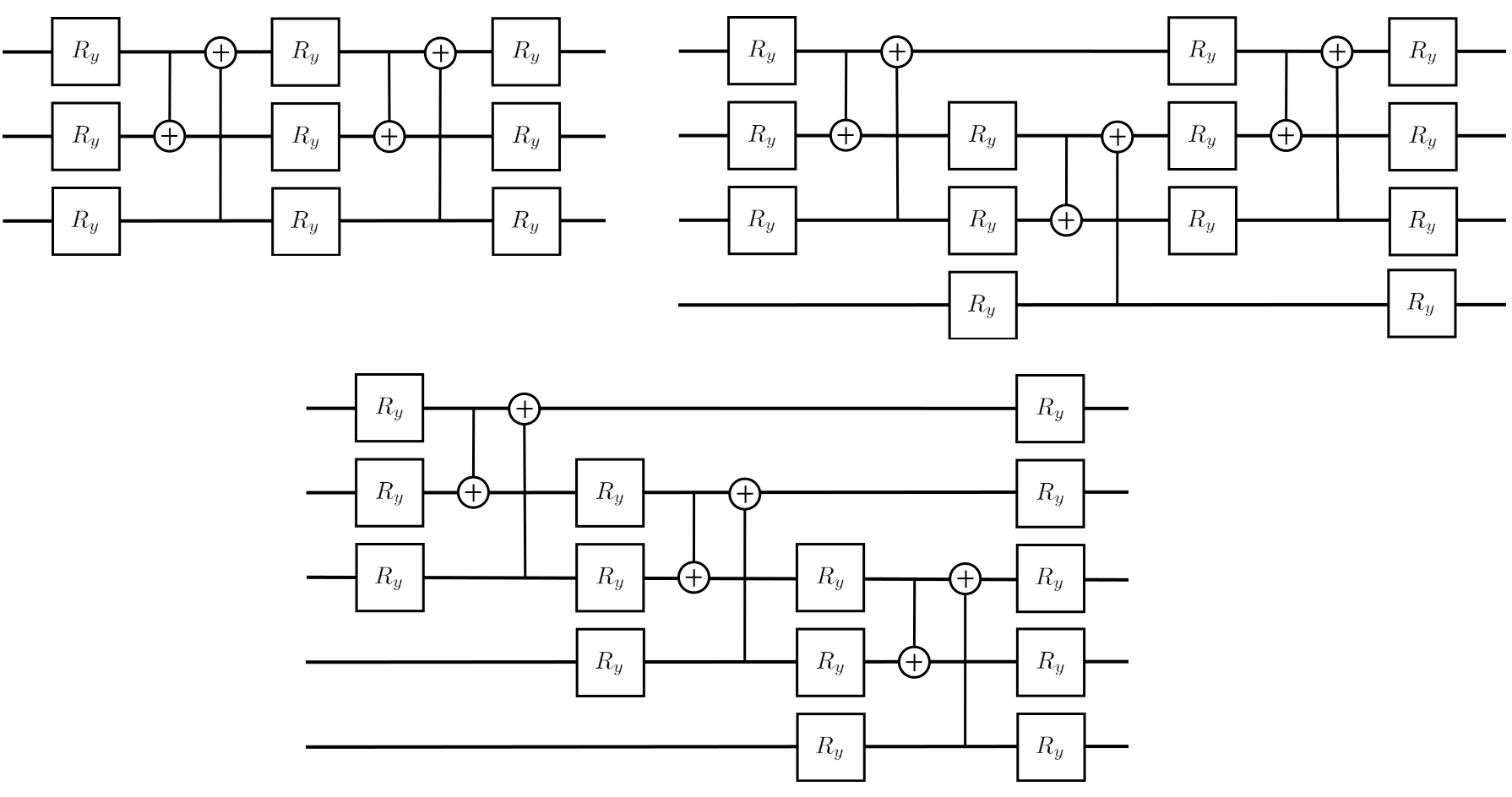}
\caption{The ans\"atze selected for the 3, 4 and 5 qubit problem instances.}
\label{ansatze}
\end{figure}

The ans\"atze selected for each of the problem instances are shown in the figure above Fig. \ref{ansatze}.
Ansatz selection was not done with any specific backend in mind, and as such the ansatz used have been assumed to be hardware efficient. No transpilation to any specific backend connectivity was done, even for the realistic noise simulation.

\subsection{Adiabatic Ansatz Variational Quantum Linear Solver}
\label{aavqls_results}

\begin{figure}[h]
\centering
\includegraphics[width=\columnwidth]{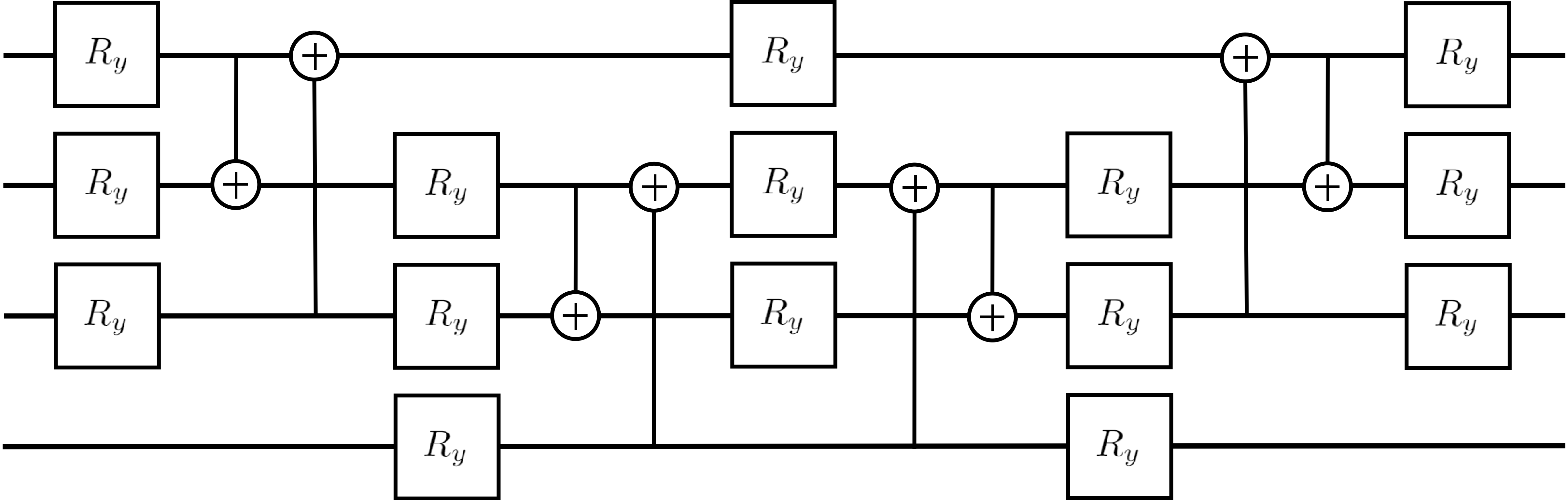}
\caption{The ansatz employed in the AAVQLS test. This ansatz can be initialized to identity with some random parameters. The circuit $U$ for creating $\ket{b}$ is appending to the end of this ansatz in accordance with the method discussed in the AAVQLS description in order to create the adiabatic ansatz.}
\label{adiabatic_ansatz}
\end{figure}

The following test of the AAVQLS algorithm compares the Adiabatic Ansatz method to a standard VQLS approach, for the three same noise levels as the test above. The same local cost function was used, and Powell's method was used as the classical optimizer \cite{powell}, due to its noise resistance \cite{paper_1}. Both the AAVQLS and the VQLS used the same ansatz circuit given in Fig. \ref{adiabatic_ansatz}. This ansatz can be initialized to identity with some non-zero random parameters, hence its use in the AAVQLS here. The circuit $U$ for creating state $\ket{b}$ is appending to the end of this ansatz in accordance with the method discussed in the AAVQLS description, to create the full adiabatic ansatz.

The example five qubit problem tested was given by,

\begin{center}
$\mathbf{A}_{4} = Z_{1} + 0.15\cdot Z_{3} + 0.5\cdot Z_{4}$.
\end{center}

The AAVQLS approach was split into two trials. One using 10 steps and the other using 20 steps, denoted in the results with the suffix `1' and `2' respectively. These were compared to a standard VQLS implementation. The same number of overall function evaluations were allowed for all approaches. 20 runs for each approach and noise-level were conducted.

\subsection{Evolutionary Ansatz Variational Quantum Linear Solver}
\label{eavqls_results}

 The EAVQLS method was compared to three standard VQLS methods. Again the local cost function was used for both the EAVQLS and the VQLS, and the 4 Qubit ansatz in Fig. \ref{ansatze} was used for the VQLS comparison. The three standard VQLS methods differ by their use of three different classical optimizers: Cons\-trained Optimization by Linear Approximation (COBYLA), BFGS and SPSA. The EAVQLS used the COBYLA \cite{cobyla} optimizer, as the simulation was done on a noise-free statevector simulator, and COBYLA has a very quick convergence rate on a noise-free simulation \cite{paper_1}. The test problem consisted of four qubits with $A$ given by,
\begin{center}
$\mathbf{A}_{5} = Z_{1} + 0.15\cdot X_{2}Z_{3} + 0.5\cdot H_{4}$.
\end{center}
20 Runs of the EAVQLS genetic algorithm were run, with a population of size 20 and 20 generations. The genetic hyper-parameters, topological search, parameter search and removal, were set to the values of 0.7, 0.2 and 0.4 respective\-ly. Each VQLS instance was run 100 times, and the top 20 best runs were selected in the comparison. This is done to give a fair comparison to the resource intense EAVQLS method and even 100 Runs of a standard VQLS algorithm requires less quantum resources than the 20 EAVQLS Runs.

\subsection{Classical Combination of Quantum States}
\label{cqs_results}

\begin{figure}
\centering
\includegraphics[width=\columnwidth]{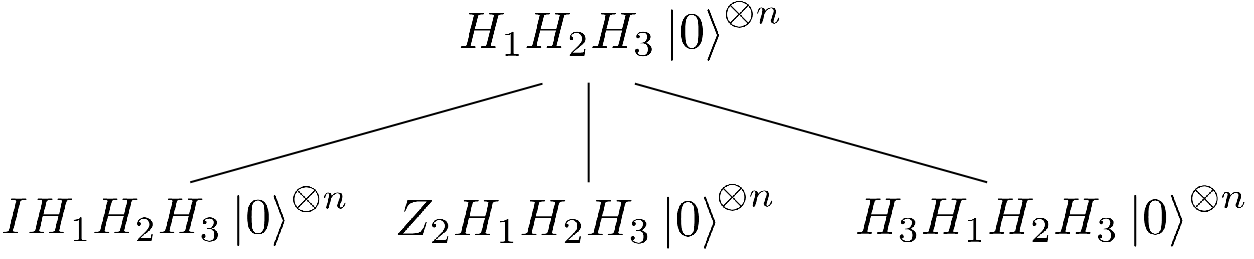}
\caption{CQS algorithm ansatz tree expansion used to solve the CQS test problem on the real quantum machine.}
\label{cqs_ansatz}
\end{figure}

The CQS method is the only non-variational approach tested. The aim of this test was to see how accurately the real quantum machine could approximate the solution given the nodes. This test follows the standard implementation of the CQS algorithm, however no ansatz tree expansion was conducted on the real device. Instead the ansatz tree was pre-expanded with the nodes as seen in Fig. \ref{cqs_ansatz}, and then the solution was approximated on the real device. 

The three qubit test example used is given as follows,
\begin{center}
$\mathbf{A}_{6} = \mathds{1} + 0.25\cdot Z_{2} + 0.175\cdot H_{3}$.
\end{center}
This example was selected as a non-trivial problem that suited the topology of the real backend selected (the IBMQX2 quantum device). The number of shots per Hadamard test was set to 245760 (being 30 repetitions of the max 8192 shots per run).

\subsection{Logical Ansatz Variational Quantum Linear Solver}
\label{logical_ansatz_results}

The logical ansatz used the same cost function as the CQS method as detailed above. Both COBYLA and Powell's method were tested as classical optimizers, and both a noise-free Statevector simulation, and a shot-noise Qasm simula\-tion were tested.
The logical ansatz was tested on the five qubit problem given by,
\begin{center}
$\mathbf{A}_{7} = H_{1} + 0.25\cdot Z_{3} + 0.5\cdot H_{4} + 0.5\cdot Z_{5}$.
\end{center}
The individual ansatze making up the logical ansatz were generated randomly. Each logical ansatz was made up of five shallow physical ansatze. Each ansatz consisting of 3 layers of gates, taken from the gate set $\{\mathbb{I}_{2}, R_y, \wedge_{1}X\}$. Two different approaches to training were tested, each denoted in the results by the suffix `1' or `2', for the first and second approach respectively. These two approaches are detailed as Method 1 and Method 2 in the Logical Ansatz section (Section. \ref{la_section}).  
Twenty runs for each approach were performed in order to obtain the results.

\section{Results and Discussion}

Below are the results for the above mentioned experiments. Please note that the cost values achieved by the VQLS, AAVQLS and EAVQLS are not directly comparable to the cost values achieved by the CQS and LAVQLS due to a differing cost function.

\subsection{Variational Quantum Linear Solver}

\begin{figure}[h]
\centering
\includegraphics[width=\columnwidth]{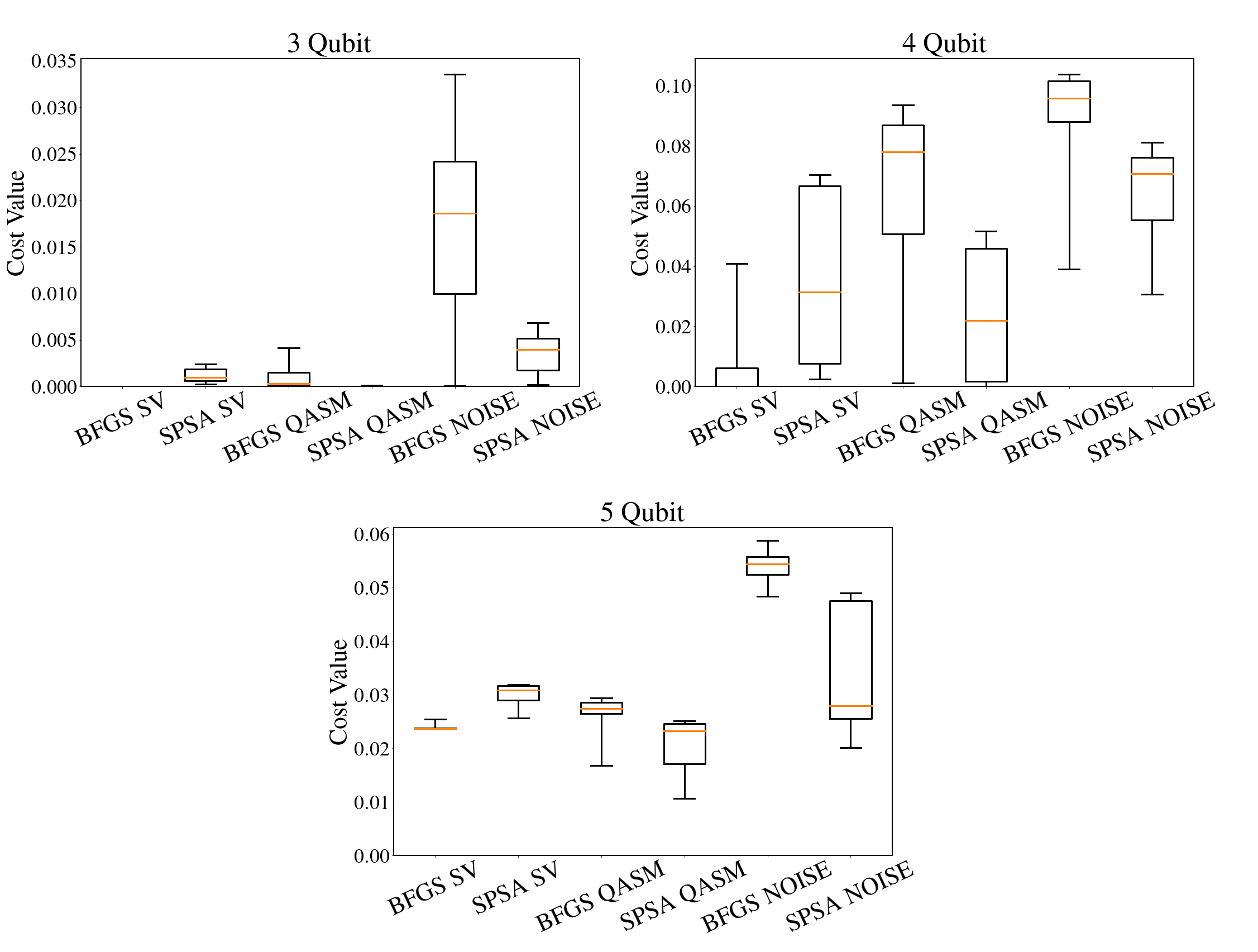}
\caption{Termination Value Box plots: These box plots capture the final value at the termination of top 50 attempts of the VQLS algorithm for both the SPSA and BFGS optimizers, with three levels of noise in simulation, for 3 problems. The standard VQLS performance is greatly affected by the presence of noise in the quantum simulation.}
\label{vqls_boxplot}
\end{figure}

\begin{figure}[h]
\centering
\includegraphics[width=\columnwidth]{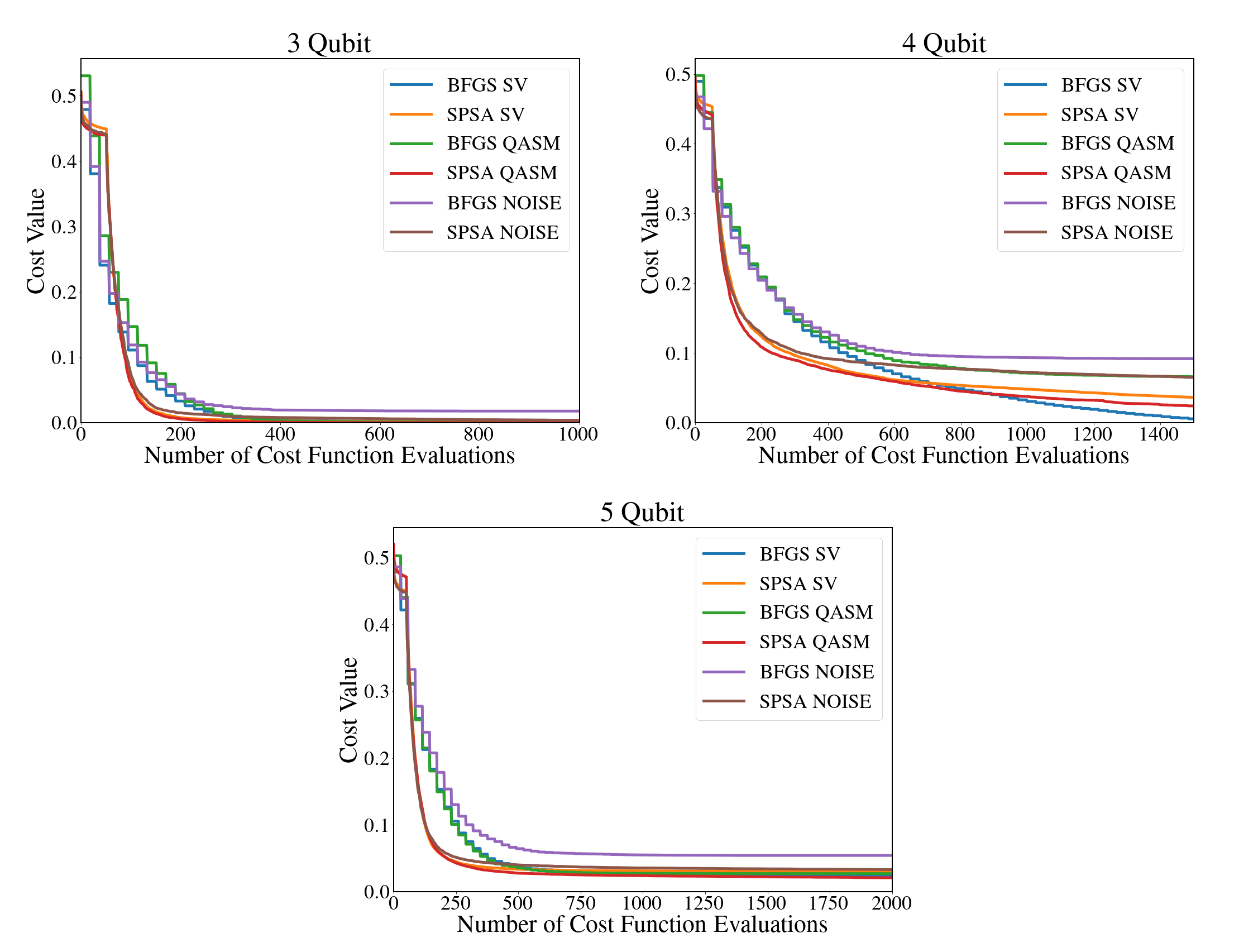}
\caption{Average Optimization Convergence: These line graphs capture the convergence of the top 50 attempts of the VQLS algorithm for both the SPSA and BFGS optimizers, with three levels of noise in simulation, for 3 problems. (BFGS convergence appears stepped as gradient function calls require multiple cost function evaluations). Both optimizers converge relatively quickly, close their final values, irrespective of the noise present.}
\label{vqls_linegraph}
\end{figure}
%\FloatBarrier
The box plot Fig. \ref{vqls_boxplot} and line graph Fig. \ref{vqls_linegraph} show the range of the termination values and the average rate of convergence respectively, for each problem instance, noise-level and classical optimizer used.

Fig. \ref{vqls_boxplot} gives an idea of the difficulty of each problem instance, and also gives an idea as to the overall affect of noise on the optimi\-zation process. The 4 qubit problem instance appears to have been the most difficult, next being the 5 qubit instance, with the 3 qubit instance being the simplest to solve.

The VQLS algorithm performs very well in a noise-free state vector simulat\-ion, with the gradient-based BFGS undoubtedly performing the best outright. The inclusion of shot-noise alone does not appear to greatly disturb the optimi\-zation process much, however BFGS is much more affected by the shot-noise than the gradient-free SPSA. SPSA very much outperforms BFGS in the presence of noise. The realistic noise levels appear to greatly affect the optimi\-zation process, and while again, SPSA is less affected than BFGS, both are heavily set back. These trends seen between the different noise levels and classical optimizers appear to hold regardless of the problem instances apparent difficulty.

Fig. \ref{vqls_linegraph} shows the average convergence rate of the top 50 attempts, for each noise-level, classical optimizer and problem instance. In all, SPSA converges faster than BFGS regardless of noise-level, and the more difficult the problem, the slower the rate of convergence.

\subsection{Adiabatic Ansatz Variational Quantum Linear Solver}

\begin{figure}[h]
\centering
\includegraphics[width=\columnwidth]{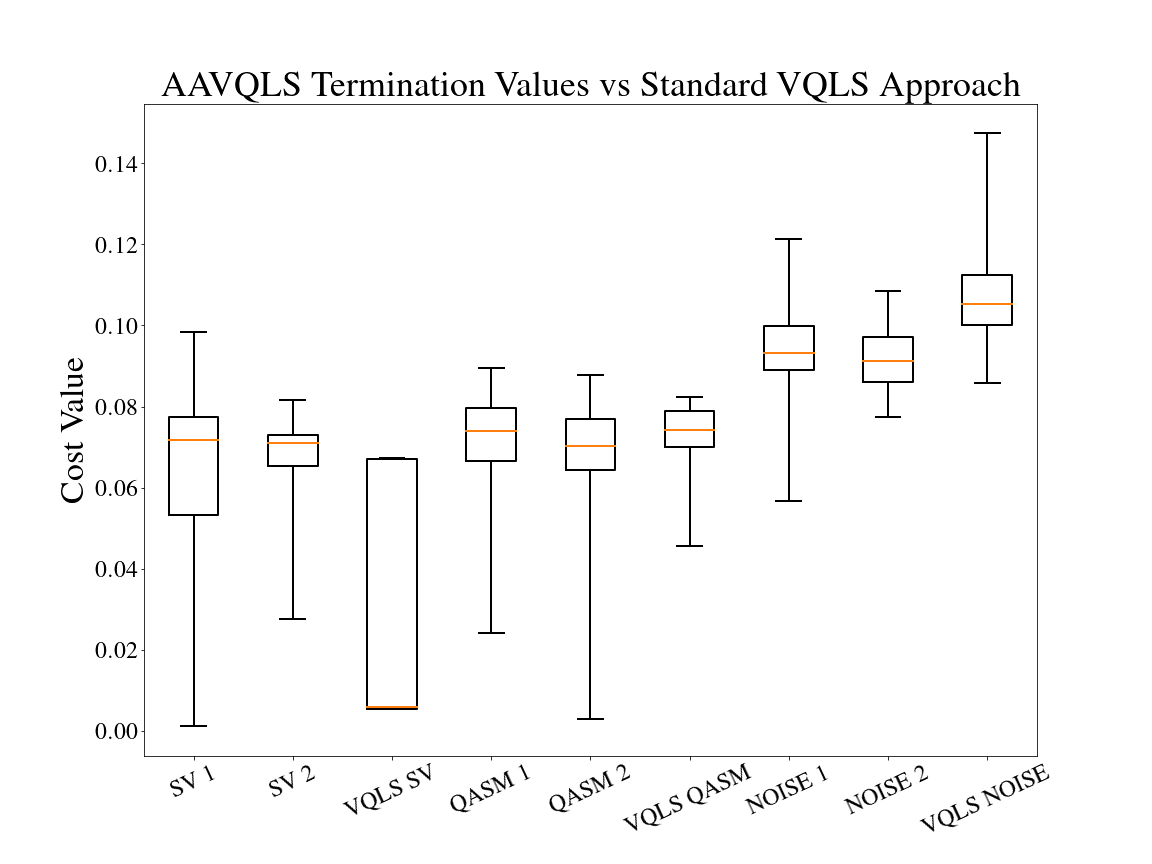}
\caption{AAVQLS Termination Values: AAVQLS termination values compared to a standard VQLS approach, for 2 different AAVQLS methods for 3 noise levels. These results appear to indicate there is not much of a significant difference between the VQLS and both AAVQLS approaches.}
\label{aavqls_boxplot}
\end{figure}
\begin{figure}[h]
\centering
\includegraphics[width=\columnwidth]{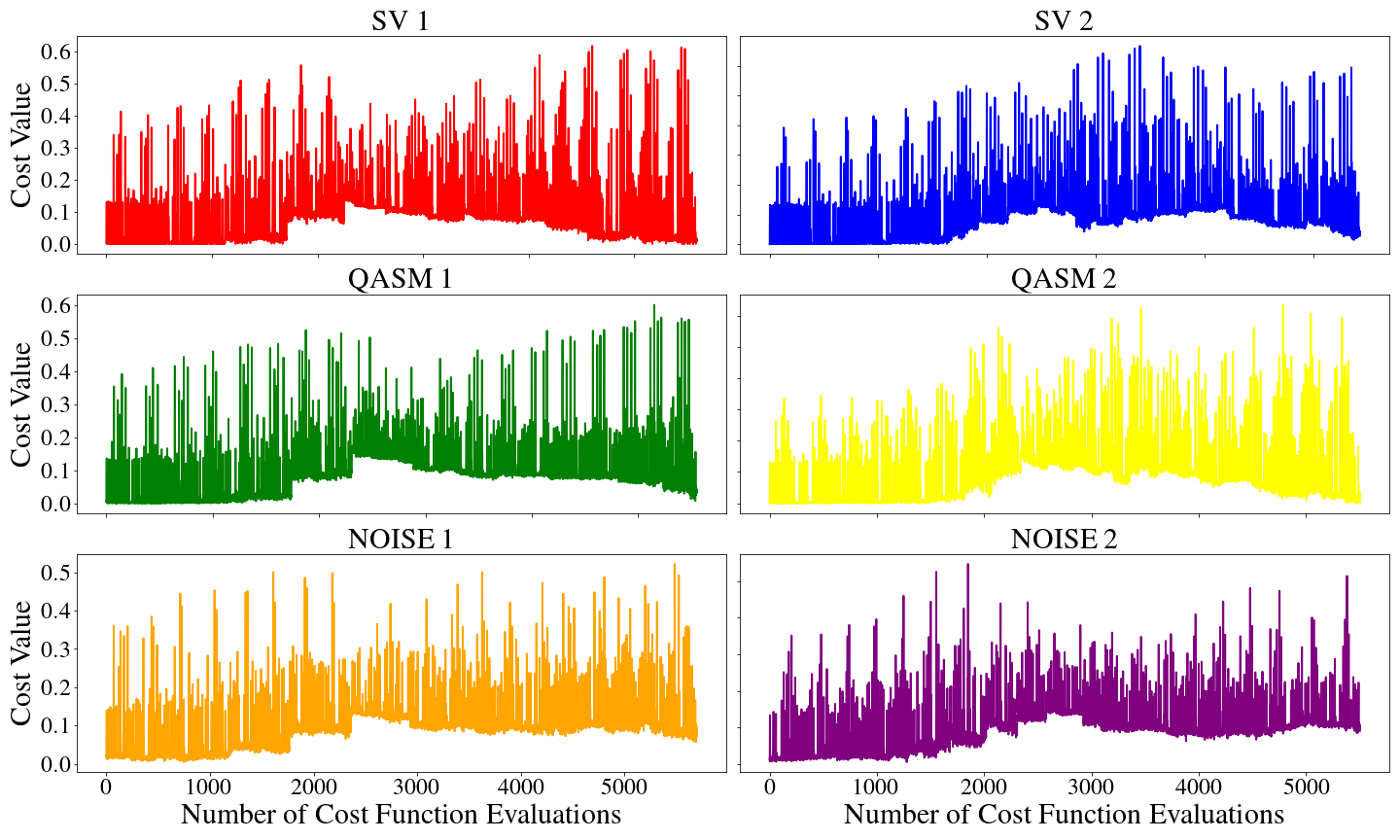}
\caption{Average Optimization Convergence: These line graphs capture the cost value measured by the AAVQLS algorithm during the optimization process, for 2 different methods, for 3 noise levels. This corresponds to how close the ansatz kept to the ground state of the Hamiltonian during the optimization.}
\label{aavqls_linegraph}
\end{figure}

The box plot Fig. \ref{aavqls_boxplot} shows the range of the final termination values achieved by the respective methods for the respective levels of noise. 

The line graph Fig. \ref{aavqls_linegraph} shows an average value of the cost function along the adiabatic optimization path for the best performing run of both AAVQLS 1 and 2 for each noise level. 

The results captured in Fig. \ref{aavqls_boxplot} appear to indicate there is not much of a significant difference between the VQLS and both AAVQLS approaches. However, the state vector simulation clearly favours the standard VQLS appro\-ach, while the both AAVQLS approaches slightly outperform the VQLS in the realistic noise situation. Given that the state vector simulation is merely theoretical, there may be some merit to the AAVQLS approach. AAVQLS 2 also ever so slightly outperforms AAVQLS 1 in the noisy simulations, meaning shorter, more frequent steps may be the better approach between the two. (The AAVQLS approach was split into two trials. One using 10 steps and the other using 20 steps, denoted in the results with the suffix `1' and `2' respectively)

The line graph Fig. \ref{aavqls_linegraph} shows that all methods kept fairly close to the ground state of the Hamiltonian during the initial phase of optimization, only to move further away from the ground state during the middle of the optimization process. The Statevector and Qasm simulations of both methods managed to move close to the ground state near the end of the optimization process however the noisy simulations did not. Ideally all methods should have kept fairly close to the ground state throughout the optimization process. 

\subsection{Evolutionary Ansatz Variational Quantum Linear Solver}
\begin{figure}[h]
  \centering
  \includegraphics[width=10cm]{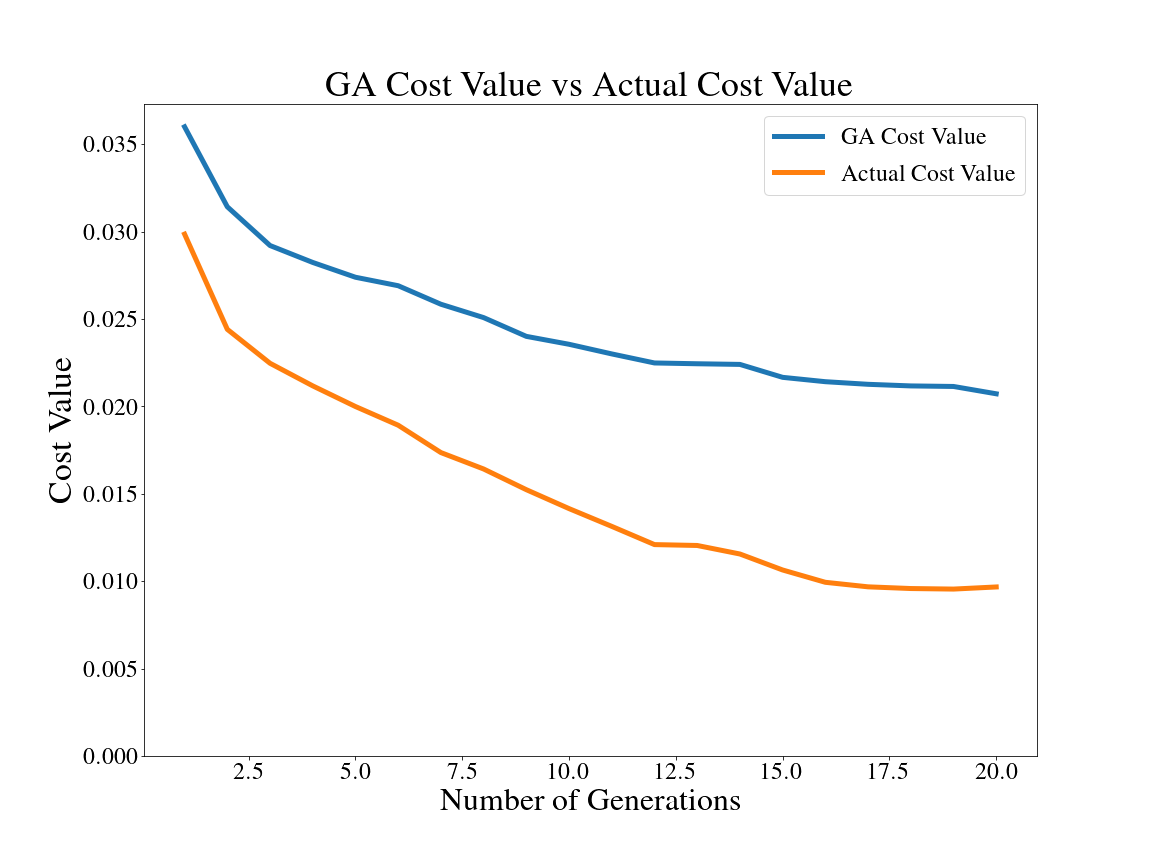}
  \caption{The best performing genome's fitness value vs the actual cost value, given by that same genome's ansatz by the local cost function, averaged over the 20 runs, for the 20 generations. Both the GA cost value and Actual cost value continue to decrease as the algorithm runs.}
  \label{eavqls_linegraph}
\end{figure}
\begin{figure}[h]
  \centering
  \includegraphics[width=\columnwidth]{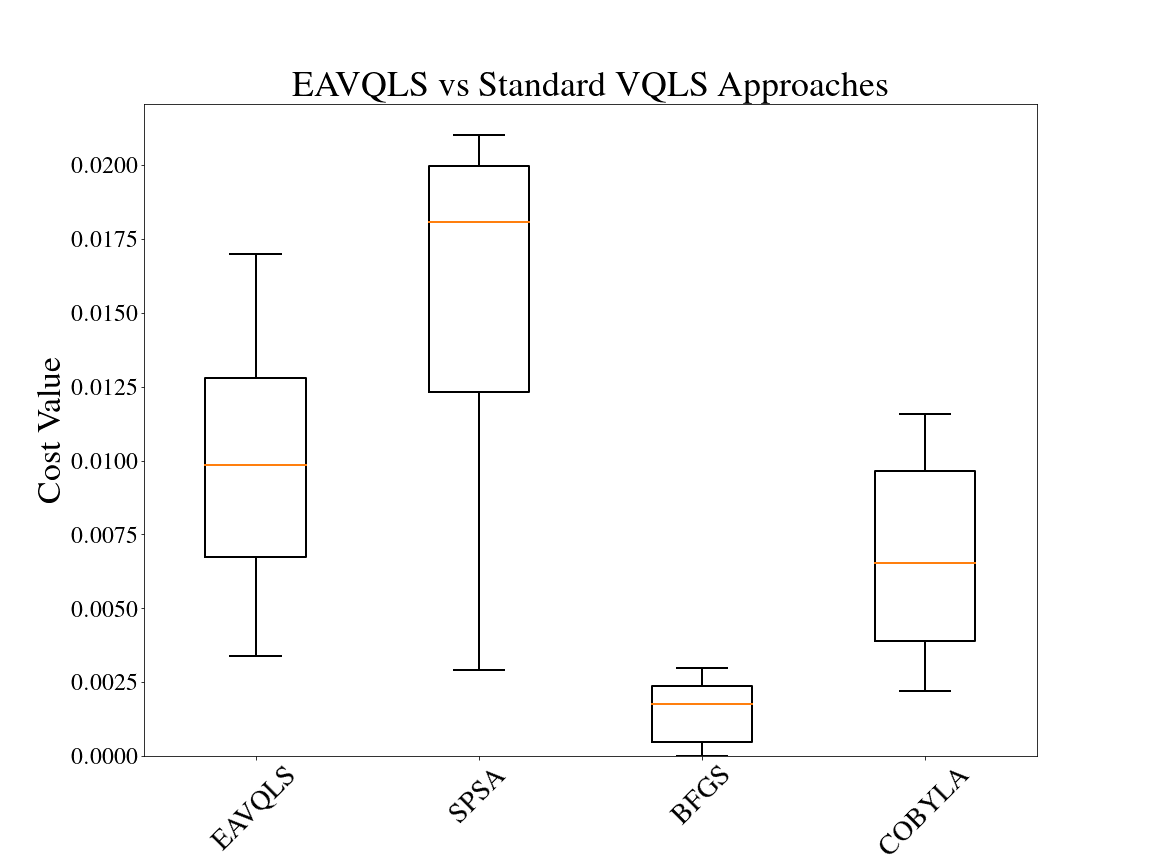}
  \caption{Cost value achieved by the EAVQLS algorithm compared to standard VQLS approaches of using SPSA, BFGS and COBYLA minimizers. The EAVQLS appears to have been outperformed by the standard VQLS using the BFGS and COBYLA optimizers.}
  \label{eavqls_boxplot}
\end{figure}
%\FloatBarrier

The line graph Fig. \ref{eavqls_linegraph}, shows the difference between the best genome's fitness value (discussed in EAVQLS section) and the actual local cost function value per generation averaged across the 20 runs.

The box plot Fig. \ref{eavqls_boxplot} shows the difference between the final value of the local cost function for the 20 EAVQLS runs and the 20 best VQLS runs per each approach. 

\subsubsection{Classical Combination of Quantum States}

\begin{figure}
  \centering
  \includegraphics[width=\columnwidth]{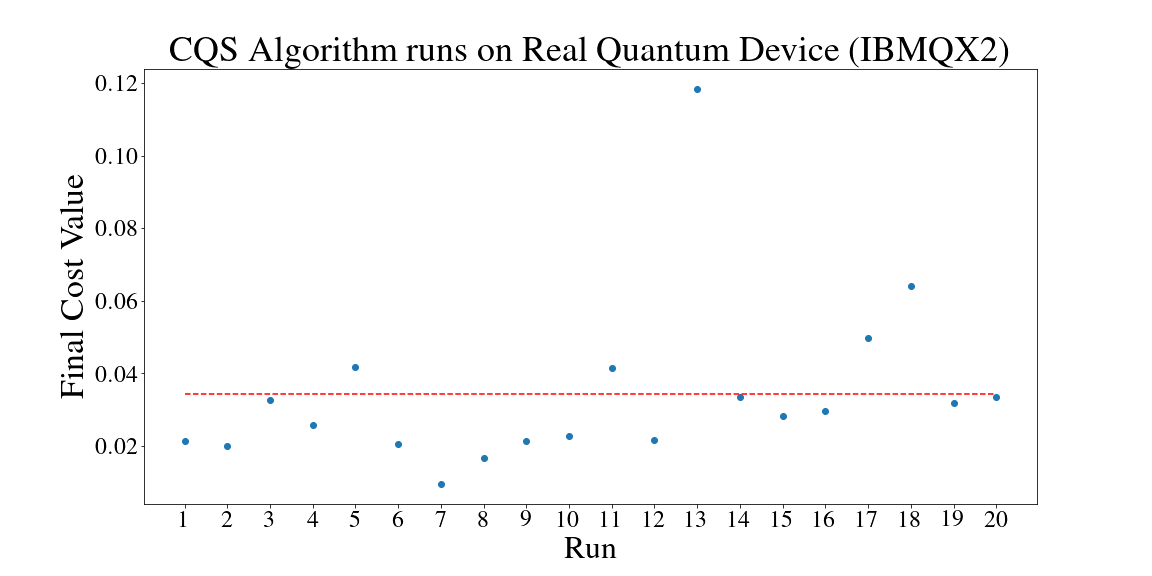}
  \caption{20 runs of the CQS Algorithm on IBMQX2 machine, with the average cost indicated in red.}
  \label{cqs_runs}
\end{figure}
%\FloatBarrier

Relatively good results achieved on the real machine Fig. \ref{cqs_runs}, given the noise levels of current NISQ devices, however the problem was specifically tailored to suite the connectivity of the selected backend. The optimal cost value achievable using the nodes in the ansatz tree given is equal to approximately $0.00324$. %$0.003236466330187704$

% \subsubsection{Results}
\begin{figure}[h]
  \centering
  \includegraphics[width=\columnwidth]{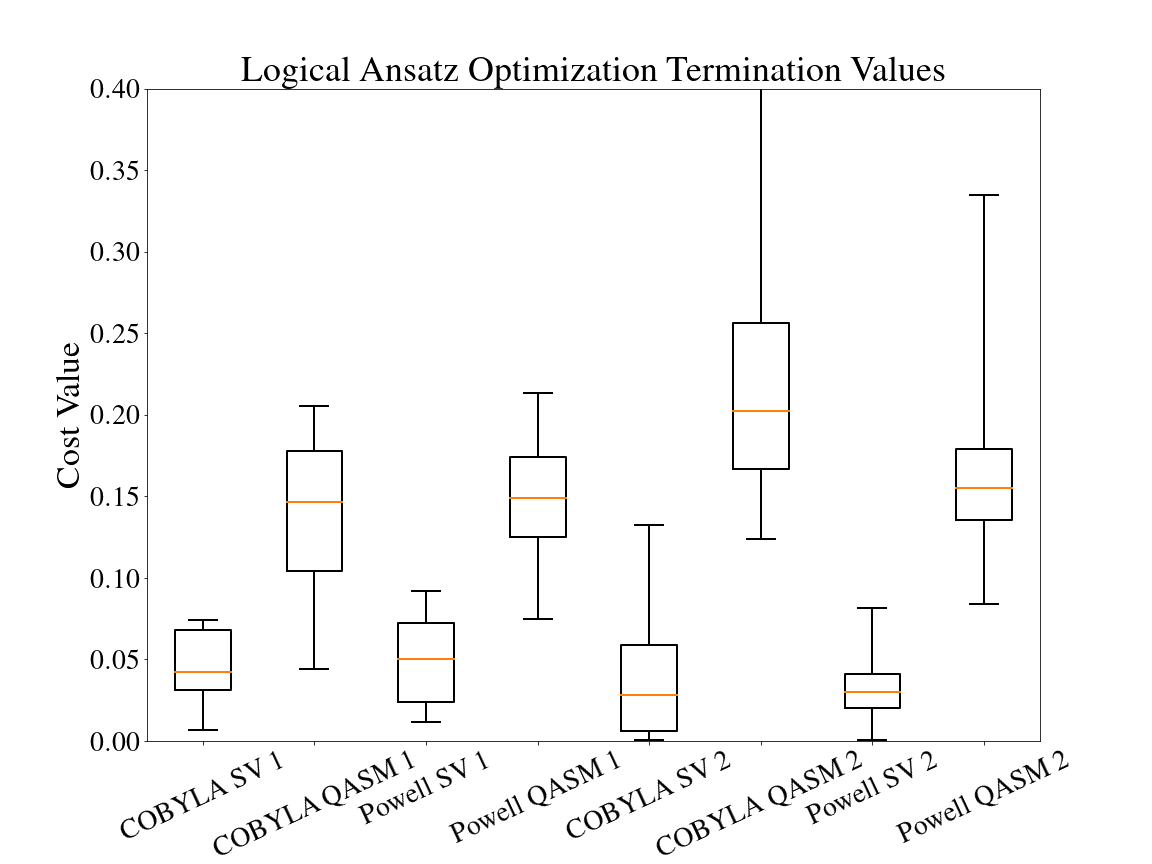}
  \caption{Termination Values achieved by the logical ansatz, for 2 methods, 2 noise levels and 2 different optimizers. Both training methods 1 and 2 performed similarly well.}
  \label{la_boxplot}
\end{figure}
%\FloatBarrier
Fig:\ref{la_boxplot} shows the spread of the final cost values of the logical ansatz approach. It is clear that the noise introduced by the Qasm simulator greatly affects the performance of the Logical Ansatz VQLS. The effect of noise may be increased when using a logical ansatz as multiple different hadamard test runs are needed, one for each pairing of the different physical ansatze, each adding some potential for error.

\section{Conclusions}
\label{conclusion}
The success of the standard VQLS approach appears to be greatly dependant on the noise levels on the quantum device. The cost values achieved remain fairly consistent after the addition of shot-noise, and decline substantially when realistic levels of noise are added to the simulated quantum device. 

The AAVQLS approach presents an ansatz optimization strategy that in theory could keep the ansatz near the ground state of the system's Hamiltonian, allowing a low cost value to be easily found. Considering the best runs recorded in Fig. \ref{aavqls_linegraph}, this trend is observed during the first part of the optimization where at all levels of noise, the system remained close the the ground state of the Hamiltonian. This trend faded just before midway through the optimization process, where all simulations, regardless of the level of noise, moved away from the ground state. This presents a particular flaw in this approach, whereby the system can leave the ground state. One possible explanation for this is that the optimization process had a step size that was too large (the evolution of the Hamiltonian was too fast), or the ground state was not contained in the Hilbert space spanned by the particular ansatz used. The latter issue may be avoided by evolving the initial Hamiltonian to the final Hamiltonian along a different path. In the later half of the optimization process, the Statevector and QASM simulations recovered the ground state while the realistic noise simulation did not. That the AAVQLS performs similarly to the VQLS in the QASM and Noisy simulations, as seen in Fig. \ref{aavqls_boxplot}, suggests that there may be some merit to this approach, especially because the best cost values achieved by the AAVQLS for those two simulations were quite substantially lower, even if, on average, they performed similarly.

The EAVQLS performed at around the same level as the VQLS for the specific problem instance simulated in these results. Seeing as only a statevector simulation was conducted, it is not yet known how a noisy simulation may have changed these results, as a key selling point of the EAVQLS algorithm is noise resistance, due to shorter, more problem and hardware specific ans\"{a}tze.

The CQS approach was the only approach tested on the real quantum device, in order to gauge its effectiveness on near-term quantum hardware. With 20 runs on the IBMQX2 device, the CQS approach managed to achieve some fairly low cost values and a decent average cost value. This is positive for this approach, however it is noted that the specific problem that was solved may have been quite simple, yet still non-trivial.

The LAVQLS, being an adaptation of the CQS method, appears to work well in a noise-free simulation, however the shot-noise alone heavily reduced the final cost value achieved by the method. This may be because the many hadamard tests required to evaluate the cost function amplify the noise. This is not good because a proposed feature of the LAVQLS was noise resistance due to the use of shorter individual ans\"{a}ze making up the logical ansatz. 

In this paper a few approaches to solving systems of linear equations on near-term quantum hardware have been presented. Each approach that differs from the standard VQLS approach tries to offer some advantage over the standard approach, however whether the proposed advantages of these algorithms actually apply in implementation is yet to be conclusively seen. Some potential problems with these approaches have been highlighted and it is left to a future work to investigate the realistic advantages of these approaches. It may be the case that some of these approaches offer significant advantages over the standard VQLS approach, however this is still unclear.

\section*{Acknowledgments}
\label{ACKNOWLEDGMENTS}
This work is based upon research supported by the South African Research Chair Initiative, Grant No. 64812 of the Department of Science and Innovation and the National Research Foundation of the Republic of South Africa. Support from the NICIS (National Integrated Cyber Infrastructure System) e-research grant QICSA is kindly acknowledged.

\end{document}